\documentclass[pra,aps,twocolumn,showpacs]{revtex4}
\usepackage{graphicx}
\usepackage{color}

\begin{document}

\title{Security improvement of using modified coherent state for quantum cryptography}

\author{Y. J. Lu}
\altaffiliation{currently with Photonics Industries International, Bohemia, NY} 

\author{Luobei Zhu}
\altaffiliation{Also with Carmel High School, Carmel, IN}

\author{Z. Y. Ou}\email{zou@iupui.edu}

\affiliation{Department of Physics, Indiana University-Purdue University Indianapolis \\ 402 N Blackford Street, Indianapolis, In 46202}

\date{\today}

\begin{abstract}
Weak coherent states as a photon source for quantum cryptography have limit in secure data rate and transmission distance because of the presence of multi-photon events and loss in transmission line. Two-photon events in a coherent state can be taken out by a two-photon interference scheme. We investigate the security issue of utilizing this modified coherent state in quantum cryptography. A 4 dB improvement in secure data rate or a nearly two-fold increase in transmission distance over the coherent state are found. With a recently proposed and improved encoding strategy, further improvement is possible.
\end{abstract}

\maketitle


\section{Introduction}
The ultimate security of quantum cryptography stems from the non-cloning theorem\cite{woo} of quantum mechanics, which is applied to a single quantum system. In other words, quantum
information can only be shared between two parties, the one (Alice) who creates it and the one (Bob) who receives it. Therefore, in the implementation in optical communication, the ideal source is a single photon source with information carried by individual photons. So far a weak coherent state from a laser is closest to a single photon source for quantum cryptography. The experimental realization\cite{BB92} of the first proposed quantum cryptography protocol, known as BB84 protocal\cite{bb84}, involved a heavily attenuated thermal source (which is worse than a laser in the aspect of multi-photon events). However, the existence of multiple photon events for a weak coherent state, even though very rare, poses serious problem for the security of the protocol. An eavesdropper (Eve) can in principle use the so-called photon number splitting attack to tap the information without being noticed.  Recent analysis by L\"utkenhaus\cite{lut} on the security of
quantum cryptography by a weak coherent state puts limit on the data rate and transmission distance for secure key distribution. 

Significant progress has been made in producing single photon states\cite{demartini,kitson,brunel,fleury,kurtsiefer,brouri,kim,michler,yama}. But so far these sources are fluorescence-based and have small efficiency. Furthermore, because the emission depends on energy  structure, it has a fixed wavelength and a finite time response for high repetition rate. So they are limited for practical applications. Recently a new scheme was proposed\cite{singh} and realized\cite{lu-ou} for producing a photon source with same single photon rate as a coherent state but less two-photon events. The scheme relies on a two-photon quantum interference phenomenon to reduce and eventually cancel the two-photon events from a coherent state. Although it cannot take out all the multi-photon events, it does take out the most dominating multi-photon events --
two-photon events. This modified coherent state (MCS) has the advantages of high data rate, well defined direction and wave-length independence for practical application in quantum cryptography. 

A new quantum cryptography protocol(SARG04) was recently proposed\cite{gisin} against the photon number splitting attack for weak coherent state. Such a scheme relies on a smarter encoding scheme that is based on nonorthogonal states and requires at least two or more photons for Eve to recover the data. Thus Eve needs at least three photons to use the photon number splitting attack. If we use this protocol on the modified coherent state, we will make a three-photon interference effect to cancel the three-photon term so that Eve can only rely on the 4-photon or higher order terms to apply the photon number splitting attack. So the modified coherent state still has advantage over a weak coherent state in this new protocol.

Because of the existence of the multi-photon terms, the modified coherent
state has similar security problems as the coherent state. But the influence of
those higher order terms will be smaller.  The security problem of the coherent state was
investigated by L{\" u}tkenhaus \cite{lut} and Brassard et al\cite{bras}.   In the following, we will apply the same line of argument of Ref.\cite{lut} to look for the optimum secure  data rate and transmission distance for a given transmission loss and dark counts of the detectors for the modified coherent state in both BB84 protocol in Ref.\cite{bb84} and the new SARG04 protocol in Ref.\cite{gisin}. To do that, we need to find the probability $P_m$ for multi-photon event and the probability $P_s$ to detect any photon in such a state. We will start in the following with the description of the modified coherent state.

\section{Generation of the modified coherent state (MCS)}
The modified coherent state is just a coherent state without the two-photon or three-photon terms.
It can be generated by mixing a coherent state with a two-photon state from parametric
down-conversion. There are two types of mixing. The first one is to inject the coherent state into
an optical parametric amplifier with a small gain. This is the scheme first proposed by Stoler to
produce light source with sub-Poisonian statistics. The second one mixes the two fields with a
beamsplitter. In practice the first method is more direct and simpler to implement. In the
following calculation, the results are more or less same. We will consider only the first method.

For simplicity of argument, we will only treat single mode case here. In reality, this corresponds
to, for example, a single temporal mode of pulses of light. We will concentrate on the
probabilities of detecting photons in each pulse. The data rate discussed below
will be the probability per pulse. In CW case, we will consider the average photon number ($ = |\alpha|^2$ for coherent state, see
below) rather than the intensity as our parameter. So the detection time  does not appear as a
separate parameter but is included in the average photon number. However, we do require the
detection time be much smaller than the coherence time of the field to ensure the single mode
approximation.

In an optical parametric amplifier, the interaction Hamiltonian is given by 
\begin{eqnarray}
\hat H = j \hbar \xi^*  \hat a^2/2 + h.c.,
\end{eqnarray}
where $\xi$ is proportional to the amplitude of the pump field and the nonlinear
conversion coefficient of the nonlinear medium. 

With the injection of a coherent state, the initial state is simply a coherent state
$|\alpha\rangle$ and the output state becomes
\begin{eqnarray}
|\Psi\rangle_{out} = \hat {\cal U} |\alpha\rangle
\end{eqnarray}
with
\begin{eqnarray}
\hat {\cal U} =  \exp {1\over 2} (\zeta^*  \hat a^2  - \zeta  \hat a^{\dagger 2}).
\end{eqnarray}
$\zeta$ is proportional to $\xi$.
As a matter of fact, the above state is simply the two-photon coherent state or the squeezed
coherent state first discussed by Yuen\cite{yuen}. In order to find the probability of photon
number, we expand the above state in the photon number state base and obtain from Ref.\cite{yuen,MW}
\begin{eqnarray}
|\Psi\rangle_{out} = \hat {\cal U}|\alpha\rangle = \sum_{n=0}^{\infty} C_n|n\rangle,
\end{eqnarray}
with
\begin{eqnarray}
C_n = {1\over \sqrt{n! \mu}} \Big ({\nu\over 2\mu}\Big)^{n\over 2}\exp\Big({\nu^*\over 2\mu}\alpha^2
- {|\alpha|^2\over 2}\Big) H_n\Big({\alpha \over \sqrt{2\mu\nu}}\Big),
\end{eqnarray}
where $H_n$ is the $n$th-order Hermite polynomial and 
\begin{eqnarray}
\mu \equiv \cosh(|\zeta|), ~\nu \equiv {\zeta\over|\zeta|} \sinh(|\zeta|),~~
{\rm or}~ \mu^2= 1+|\nu|^2.\nonumber
\end{eqnarray}
More specifically to our interest, we list
\begin{eqnarray}
C_0 &=&   {1\over \sqrt{\mu}} \exp({\nu^*\over 2\mu}\alpha^2 - {1\over 2}|\alpha|^2),\cr
C_1 &=&  C_0 {\alpha\over
\mu},\cr
C_2 &=& {C_0 \over \sqrt{2}}  \Big[
-{\nu\over \mu}+\Big({\alpha\over\mu}\Big)^2\Big],\cr
C_3 &=& {C_0 \over \sqrt{6}} \Big[
-{3\nu\alpha\over \mu^2}+ \Big({\alpha\over\mu}\Big)^3\Big],\cr
C_4 &=& {C_0 \over \sqrt{24}}\Big[
3\Big({\nu\over\mu} \Big)^2- 6{\nu\alpha^2\over \mu^3}+
\Big({\alpha\over\mu}\Big)^4\Big],\nonumber
\end{eqnarray}

From the above, we easily find the probability for multi-photon event as
\begin{eqnarray}
P_m = 1 - |C_0|^2 - |C_1|^2.
\end{eqnarray}
This quantity is minimized when 
\begin{eqnarray}
\alpha^2 = \mu\nu
\end{eqnarray}
with a minimum value of
\begin{eqnarray}
P_m = 1 - {1\over\mu}\Big(1+{|\nu|\over\mu}\Big) e^{-(\mu-|\nu|)|\nu|}.
\end{eqnarray}
Notice that the two-photon probability $P_2 = |C_2|^2$ is  zero under this condition. This
results from a two-photon interference effect between the coherent state and the spontaneous
parametric down-conversion, i.e., interference between the two-photon amplitude $\mu\nu/\sqrt{2}$ of
parametric down-conversion and the two-photon amplitude $\alpha^2/\sqrt{2}$ of a coherent state. 

For SARG04 protocol, we need to minimize the probability for events of three or more photons. This
probability is given by
\begin{eqnarray}
P_m^{\prime} = 1 - |C_0|^2 - |C_1|^2- |C_2|^2,
\end{eqnarray}
which is minimized when
\begin{eqnarray}
\alpha^2 = 3\mu\nu
\end{eqnarray}
with a minimum value of 
\begin{eqnarray}
P_m^{\prime} = 1 - {1\over\mu}\Big(1+{|\nu|\over\mu}\Big)
\Big(1+{2|\nu|\over\mu}\Big) e^{-3(\mu-|\nu|)|\nu|}.
\end{eqnarray}
Similar to the two-photon case, when condition in Eq.(10) is satisfied, three-photon probability
$P_3 = |C_3|^2$ is zero as a result of three-photon interference. It is not easy to understand how
a three-photon effect can arise from parametric down-conversion with only even number of photons.
What happens is that a two-photon event from parametric down conversion combines with a single
photon event in coherent state to form a three-photon event with an amplitude of
$\sqrt{3/2}\mu\nu\alpha$. This amplitude interferes with another three-photon event directly from
the coherent state with an amplitude of $\alpha^3/\sqrt{6}$. Complete cancellation occurs when the
two amplitudes are equal leading to Eq.(10).

To evaluate the photon detection probability $P_s$, we notice that in practice, photodetection
usually has non-unit efficiency. This will reduce the signal level. We need to calculate the photon
detection probability under this non-ideal condition. The non-unit efficiency is modelled as a
beamsplitter of transmissivity
$\eta$. We may start with the photon number probability for single mode case as
\cite{MW}
\begin{eqnarray}
P_n = \Big\langle : {(\hat a'^{\dagger}\hat a')^n\over n!}\exp (-\hat a'^{\dagger}\hat a')
:\Big\rangle,\nonumber
\end{eqnarray}
where $: :$ denotes normal ordering and $\hat a' = \sqrt{\eta} \hat a + \sqrt{1-\eta}\hat a_0$ with
$\hat a_0$ in the vacuum. More specifically for the case of no photon:
\begin{eqnarray}
P_0 = \langle : \exp (-\hat a'^{\dagger}\hat a'):\rangle.\nonumber
\end{eqnarray}
Then a simple calculation leads to
\begin{eqnarray}
P_0 = \langle : \exp (-\eta \hat a^{\dagger}\hat a) :\rangle,
\end{eqnarray}
where the average is over the state in Eq.(2). Then the photon detection probability is simply
\begin{eqnarray}
P_s = 1 - P_0.
\end{eqnarray}
From the appendix, we have
\begin{eqnarray}
P_0 = {\exp\Big [ -{\eta x^2(\mu-|\nu|)\over
\mu+|\nu|(1-\eta)} - {\eta y^2(\mu+|\nu|)\over
\mu-|\nu|(1-\eta)}\Big ]\over\sqrt{\mu^2-|\nu|^2(1-\eta)^2}}
\end{eqnarray}
with $ x +i y  \equiv  \alpha  e^{-i\varphi/2}$ ($e^{i\varphi} \equiv \nu/|\nu|$).

With the condition in Eq.(7) for minimum multi-photon events, we have the signal probability
\begin{eqnarray}
P_s = 1-{\exp\Big [ -\eta \mu|\nu|(\mu-|\nu|)/
(\mu+|\nu|(1-\eta)) \Big ]\over\sqrt{\mu^2-|\nu|^2(1-\eta)^2}},
\end{eqnarray}
or
\begin{eqnarray}
P_s^{\prime} = 1-{\exp\Big [ -3\eta \mu|\nu|(\mu-|\nu|)/
(\mu+|\nu|(1-\eta)) \Big ]\over\sqrt{\mu^2-|\nu|^2(1-\eta)^2}}
\end{eqnarray}
for condition in Eq.(10).

\section{Optimum Secure Transmission Rate}
Next we evaluate the secure transmission rate based on the formulism from Ref.\cite{lut}. The
communication rate per slot (pulse) for a certain error rate $e$ is given by
\begin{equation}\label{}
    R = \frac{P_s}{2}\bigg [\rho (1 - \tau(e)) + f(e)h(e)\bigg ].
\end{equation}
In the above equation, the parameter $\rho$ is given by
\begin{equation}\label{}
    \rho = \frac{P_s - P_m}{P_s},
\end{equation}
which is the probability of detection events originating from the desirable photon events.
The compression function $\tau(e)$ is given by
\begin{equation}\label{}
    \tau(e) = \log_2 \bigg [1 + 4\frac{e}{\rho} - 4(\frac{e}{\rho})^2 \bigg ].
\end{equation}
The function $h(e)$ is the Shannon entropy function given by
\begin{equation}\label{}
    h(e) = -e \log_2 e - (1 - e) \log_2 (1 - e).
\end{equation}
The function $f(e)$ characterizes the performance of the error correction algorithm.

\begin{figure}
\centering \includegraphics[width=2.5in]{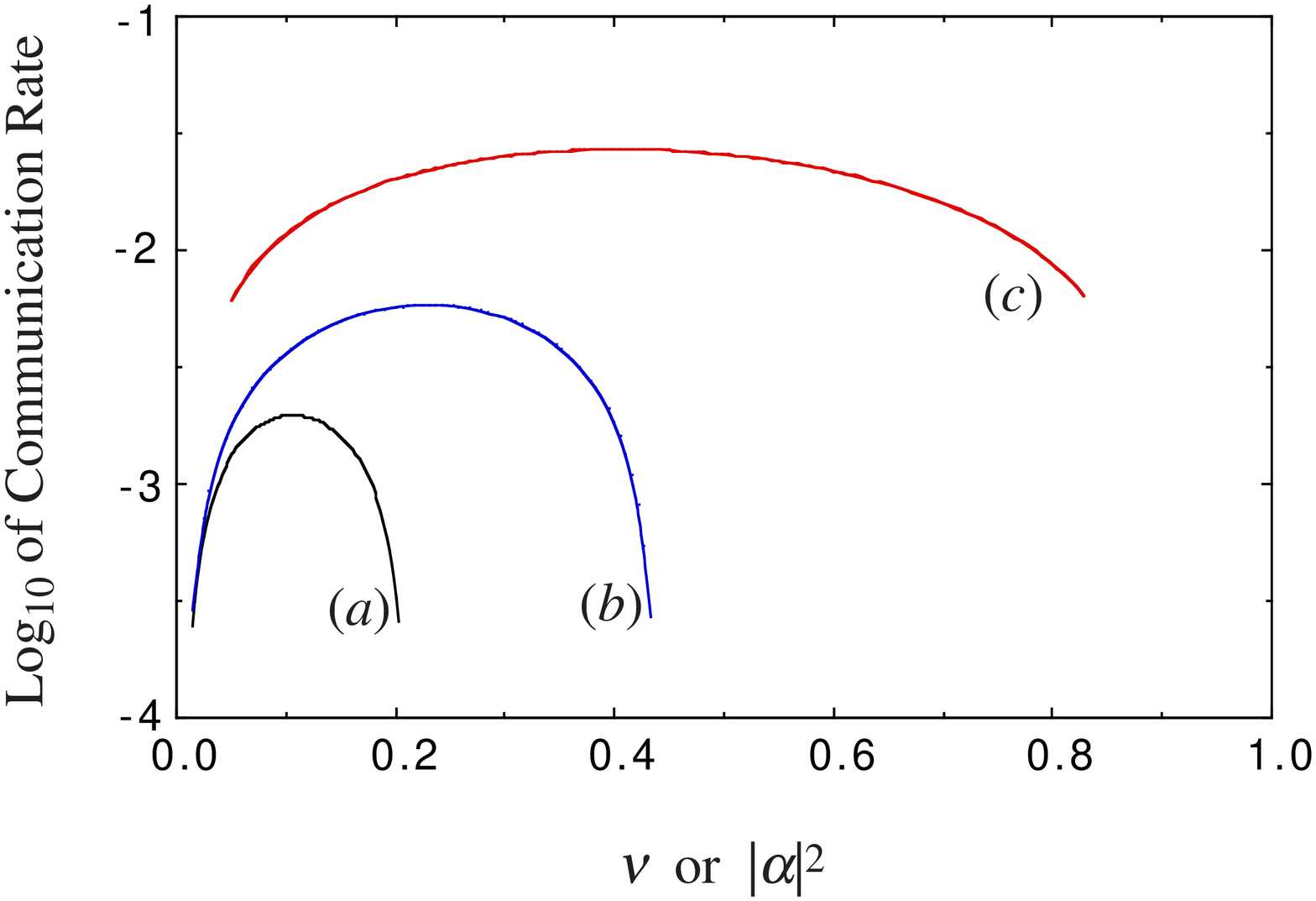}
\caption{Communication rate as a function of (a) $|\alpha|^2$ for BB84 with coherent
state, (b) $\nu$ for BB84 with modified coherent state, and (c) $\nu$ for SARG04 with modified
coherent state. The overall efficiency is $\eta$ = - 9.45 dB, which  corresponds to a transmission
distance of 5 km in KTH15.}
\end{figure}

In order to simulate the communication rate, we need the values of $P_s$, $P_m$, $e$, and
$f(e)$. For BB84 protocol with the modified coherent state, $P_m$ and $P_s$ are given in Eqs.(7) and
(15), respectively. However, due to the existence of dark count of any photodetectors, Bob's
detection events can be contributed from a signal that originates from Alice's transmission, and a
dark count that originates from Bob's photodetectors. Therefore, after taking into account the
dark counts of the detectors, the signal probability $P_s$ has to be modified as
\begin{equation}\label{}
     P_s \rightarrow \bar P_s = P_s + P_d - P_s P_d,
\end{equation}
where $P_d$ is the dark count distribution and is simply equal to the dark counts per slot.
Similarly, the error rate is also resulted from the signal and the dark count, and can be
modelled by
\begin{equation}\label{}
    e = \frac{cP_s + P_d/2}{\bar P_s},
\end{equation}
where $c$ is a constant and characterizes a baseline signal error rate. Typical system
should have a value of $c$ less than $2\%$. The total detection efficiency $\eta$ can be separated
into the channel transmission and the quantum efficiency of Bob's photodetectors. For fiber
communications, the transmission in the quantum channel is given by
\begin{equation}\label{}
    \eta_c = 10^{-(al + L)/10},
\end{equation}
where $a$ is the loss coefficient of the fiber channel measured in dB/km, $l$ is the length
of the fiber, and $L$ is the loss in the receiver Bob's detection.

We evaluate the communication rate as a function of the free parameters  $\nu (\mu
=\sqrt{1+\nu^2})$ for MCS in BB84 and SARG04 or $|\alpha|^2$ for coherent state, using
parameters taken from the case KTH15 in Ref.\cite{lut}. The dark counts per slot is $P_d = 2 \times
10^{-4}$. The baseline signal error rate is set to be $c = 0.01$. The loss coefficient $a$ of the
fiber channel is 0.2 dB/km, and the receiver loss is 1 dB. The quantum efficiency of the
photodetectors is $\eta_d = 0.18$. The total detection efficiency is then $\eta = \eta_c
\eta_d$.  For MCS in BB84, $P_m$ and $P_s$ are given in Eqs.(8) and (15)  while for SARG04 protocol
with MCS, $P_m$ and $P_s$ are from Eqs.(11) and (16), respectively. Fig.1 shows the
communication rates as a function of $|\alpha|^2$ at $l$ = 5 km for the BB84 protocol with coherent
state or as a function of $\nu$ for the modified coherent state in BB84 protocol and the SARG04
protocol. As can be seen,  there is a clear maximum on the
communication rate curve as a function of the adjustable parameter $\nu$ or $|\alpha|^2$. 

Next we concentrate on the optimal rate as we vary
the transmission distance $l$, then obtain the resulting curves as shown in Fig.2.    The overall
efficiency $\eta$ is plotted in the top axis of Fig.2 for reference.  The curve (a) is a simple repeat of Ref.\cite{lut}
for a weak coherent state in BB84. Curves (b) for MCS in BB84 and (c) for MCS in SARG04 have
significant improvement over (a). At zero transmission distance, we find a 4 dB increase in secure
data rate from modified coherent state over the coherent state in the BB84. A further 6 dB increase
in secure data rate is resulted from MCS with SARG04 protocol. Note that each curve features a
cutoff transmission distance. Modified coherent state has nearly two-fold increase in transmission
distance over the coherent state in BB84. A further nearly two-fold increase in
transmission distance is found with SARG04.

\begin{figure}[tpb]
\centering \includegraphics[width=2.5in]{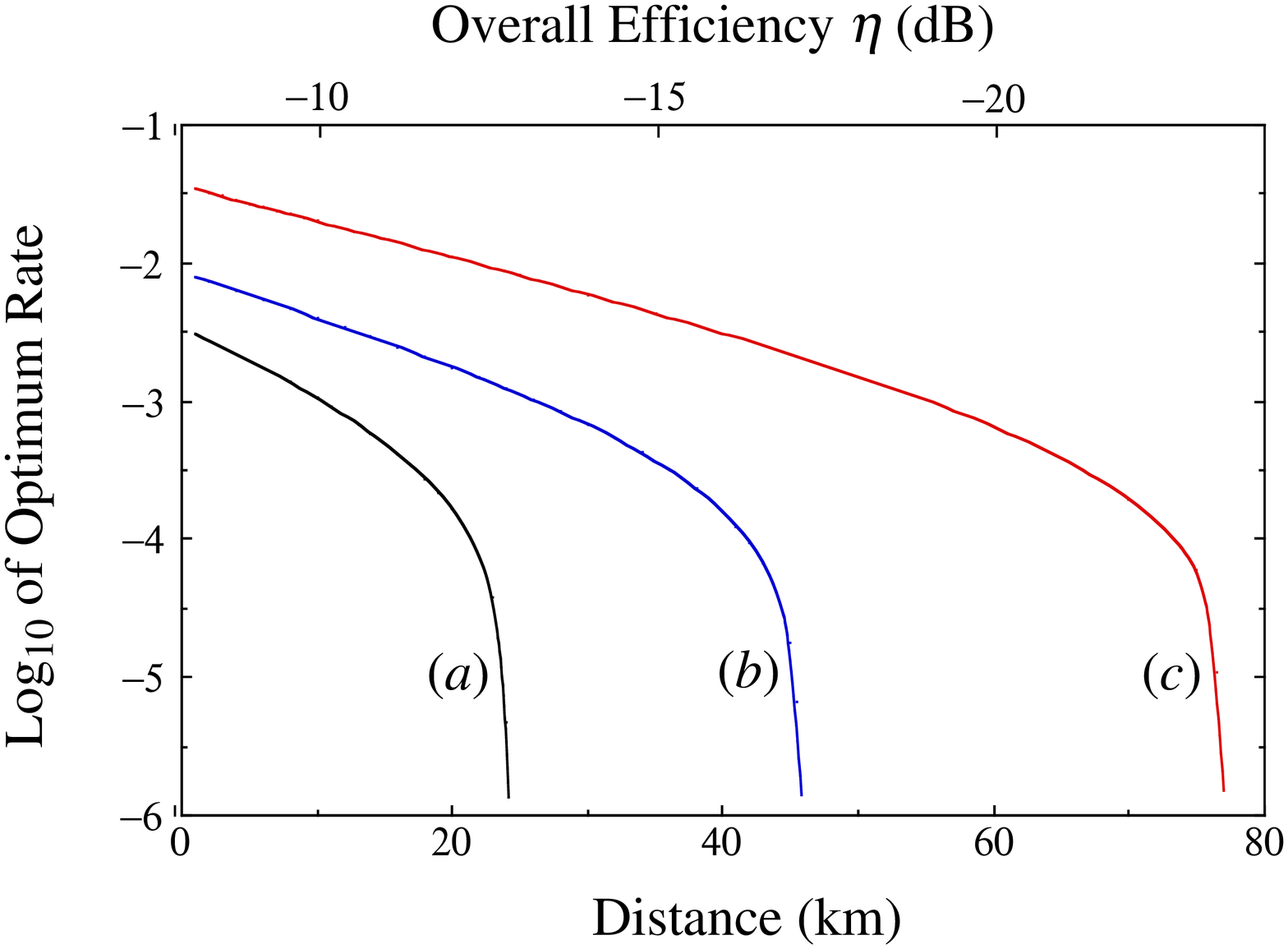}
\caption{Optimum rate as a function of transmission distance for: (a) BB84 with coherent state; (b)
BB84 with modified coherent state; (c) SARG04 with modified coherent state. }
\end{figure}

\acknowledgments

This paper is based upon work supported by the National Science Foundation under Grant No.
0245421. We wish to acknowledge the support of NSF and the Purdue Research Foundation.

\appendix

\section{Calculation of Eq.(14)}
The average in Eq.(12) is over the state in Eq.(4). 
To do this we first convert Eq.(12) in the normal ordering form into one without normal ordering
from Ref.\cite{loui}:
\begin{eqnarray}
P_0 = \langle : \exp (-\eta \hat a^{\dagger}\hat a) :\rangle
= \langle  \exp [-\ln(1-\eta) \hat a^{\dagger}\hat a) ]\rangle.
\end{eqnarray}
Next we convert Eq.(A1) into anti-normal ordering form from Ref.\cite{loui} again:
\begin{eqnarray}
P_0 
&=& e^{-\ln(1-\eta)} \langle  {\cal A}\{\exp [(1-e^{-ln(1-\eta)}) \beta^*\beta
]\}\rangle\nonumber\\
&=&{1\over 1-\eta} \langle  \alpha|\hat {\cal U^{\dagger}}{\cal A} \{\exp [-{\eta\over 1-\eta}
\beta^*\beta]\} \hat {\cal U}|\alpha \rangle.
\end{eqnarray}
Then we insert the closure relation:
\begin{eqnarray} 
{1\over\pi} \int d^2 \beta |\beta \rangle \langle  \beta | = 1
\end{eqnarray}
between $\hat b^n$ and $\hat b^{\dagger m}$ in ${\cal A}(\beta^{*m}\beta^{n})  = \hat b^n \hat
b^{\dagger m}$ in Eq.(A2) and we have
\begin{eqnarray}
P_0 =
{1\over\pi} \int d^2 \beta {1\over 1-\eta} \exp
(-{\eta\over 1-\eta} |\beta|^2)~ |\langle \beta |\hat {\cal U}|\alpha \rangle|^2.
\end{eqnarray}
From Ref.\cite{yuen,MW} we find
\begin{eqnarray}
&\langle& \beta |\hat {\cal U}|\alpha \rangle \cr
&=& {1\over\sqrt{\mu}}\exp \Big[-{|\alpha|^2+
|\beta|^2\over 2}+{\nu^*\alpha^2 -\nu\beta^{*2} +2\beta^*\alpha \over2\mu}\Big].\nonumber
\end{eqnarray}
Substituting the above expression in Eq.(A4) and with some lengthy manipulation, we arrive at the
expression in Eq.(14).

\end{document}